\documentclass[prd,aps,showpacs,nofootinbib]{revtex4}
\usepackage{amssymb}
\usepackage{graphicx}
\usepackage{amsmath}

\newcommand{\be}{\begin{equation}}
\newcommand{\ee}{\end{equation}}
\newcommand{\bq}{\begin{eqnarray}}
\newcommand{\eq}{\end{eqnarray}}

\begin{document}
\title{Lorentz violation with a universal minimum speed as foundation of de Sitter relativity} 
\author{**Cl\'audio Nassif Cruz, *Rodrigo Francisco dos Santos and A. C. Amaro de Faria Jr.}
\affiliation{\small{**CBPF: Centro Brasileiro de Pesquisas F\'isicas, R. Dr. Xavier Sigaud 150, Urca, 22290-180, Rio de Janeiro-RJ, Brazil.\\
 *UFF: Universidade Federal Fluminense, Instituto de F\'isica, Av. Gal. Milton Tavares de Souza s/nº, Gragoat\'a, 24.210-346, Niter\'oi-RJ, 
Brazil.\\
 IEAv: Instituto de Estudos Avan\c{c}ados, Rodovia dos Tamoios Km 099, 12220-000, S\~ao Jos\'e dos Campos-SP, Brazil.\\
 **cnassif@cbpf.br, *santosst@if.uff.br, antoniocarlos@ieav.cta.br}} 

\begin{abstract}
We aim to investigate the theory of Lorentz violation with an invariant minimum speed so-called Symmetrical Special Relativity 
(SSR) from the viewpoint of its metric. Thus we should explore the nature of SSR-metric in order to understand the origin of the conformal
factor that appears in the metric by deforming Minkowski metric by means of an invariant minimum speed that breaks down Lorentz symmetry.
So we are able to realize that there is a similarity between SSR and a new space with variable negative curvature ($-\infty<\mathcal R<0$)
connected to a set of infinite cosmological constants ($0<\Lambda<\infty$), working like an extended de Sitter (dS) relativity, so that
such extended dS-relativity has curvature and cosmological ``constant'' varying in the time. We obtain a scenario that is more similar to 
dS-relativity given in the approximation of a slightly negative curvature for representing the current universe having a tiny 
cosmological constant. Finally we show that the invariant minimum speed provides the foundation for understanding the kinematics origin
of the extra dimension considered in dS-relativity in order to represent the dS-length.   
      
\end{abstract}
\maketitle

\section{Introduction}

The study of de Sitter (dS) spaces presents great importance in Physics and especially in Cosmology by means of the description of the
dark energy\cite{Perlmutter} and the cosmological phenomena like the cosmic voids\cite{void1}\cite{void2} and 
cosmic inflation\cite{cosmicinflation}. 

Symmetrical Special Relativity (SSR) is a theory that has recently appeared\cite{Nassif, Nassif1, Nassif2, Nassif3, Nassif4} as a new candidate 
for describing gravitational vacuum, from where it has already been possible to calculate the value of the cosmological constant and 
also discuss the origin of the uncertainty principle\cite{Nassif3}. SSR is a kind of theory of deformed special relativity (DSR) with 
double limit of speeds, namely a maximum limit of speed $c$ (speed of light) and a minimum speed $V$. SSR presents a certain similarity
with theories that have invariant scales as proposed by Magueijo, Camelia and Smolin\cite{Maqueijo, Camelio, Smolin}.

 By making a kind of scale transformation with dependence on velocity in the metric, i.e., the factor $\Theta(v)$\cite{Nassif, Nassif2} 
acting on the Minkowski metric $\eta_{\mu\nu}$, we find SSR-metric $\mathcal G_{\mu\nu}=\Theta(v)\eta_{\mu\nu}$, which depends on velocity
($v$), thus leading us to think that it seems to be a Finsler metric of type $g(x,\dot x)$\cite{Finsler}. 
However, a deep investigation whether SSR-geommetry is a kind of Finsler or not deserves to be made elsewhere. 

 We see that SSR-theory generates an equation of state (EOS) being similar to that obtained by DS-relativities 
\cite{Sitter1, Sitter2, Sitter3}. Thus, SSR-theory provides an explanation for the effects of dark
energy (cosmological anti-gravity) by means of a new kinematics principle (an invariant minimum speed). However, unlike relativity 
theories based on dS-group, the invariant minimum speed leads to a change on Minkowski
metric by means of the introduction of the scale factor depending on velocity, i.e., $\Theta$\cite{Nassif, Nassif2}. We show that such 
a scale factor works like a conformal factor, since there is a correspondence of SSR-metric with an equivalent metric depending
on coordinate $r$ given in a dS-scenario. 

Actually, the scale factor with dependence on velocity is responsible for the introduction of a variable negative curvature 
($-\infty<\mathcal R(v)<0$), which allows us to understand the cosmological ``constant'' as being a cosmological
scalar field (parameter) varying with the age (radius) of the universe, since we show that SSR-metric behaves like a superposition of
infinite maximal spaces (dS-relativities), each one of them obtained for a certain value of $v$. Only for $v>>V$, we get
maximal spaces for representing the regime of weak anti-gravity ($\Lambda\approx 0$).

\section{The causal structure of SSR}

\begin{figure}
\begin{center}
\includegraphics[scale=0.70]{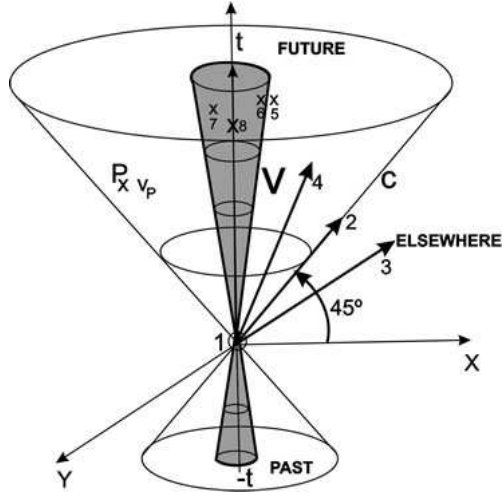}
\end{center}
\caption{The external and internal conical surfaces represent respectively the speed of light $c$ and the unattainable minimum
speed $V$, which is a definitely prohibited boundary for any particle. For a point $P$ in the world line of a particle, in the interior
of the two conical surfaces, we obtain a corresponding internal conical surface, such that we must have $V<v_p\leq c$. The $4$-interval
$S_4$ is a time-like interval. The $4$-interval $S_2$ is a light-like interval (surface of the light cone). 
The $4$-interval $S_3$ is a space-like interval (Elsewhere). The novelty in spacetime of SSR are the $4$-intervals $S_5$ (surface of
the dark cone) representing an infinitly dilated time-like interval, including the $4$-intervals $S_6$, $S_7$ and $S_8$ inside the dark 
cone for representing a new space-like region (see ref.\cite{Nassif}).}
\end{figure}

The breakdown of Lorentz symmetry for very low energies\cite{Nassif}\cite{Nassif2} generated by the presence of a
background field is due to an invariant mimimum speed $V$ and also a universal dimensionless constant $\xi$\cite{Nassif}, working
like a gravito-electromagnetic constant, namely:

\begin{equation}
\xi=\frac{V}{c}=\sqrt{\frac{Gm_{p}m_{e}}{4\pi\epsilon_0}}\frac{q_{e}}{\hbar c},
\end{equation}
$V$ being the minimum speed and $m_{p}$ and $m_{e}$ are respectively the mass of the proton and electron. Such a minimum 
speed is $V=4.5876\times 10^{-14}$ m/s. We have found $\xi=1.5302\times 10^{-22}$\cite{Nassif}. 

It was shown\cite{Nassif} that the minimum speed is connected to the cosmological constant in the following way:

\begin{equation}
V\approx\sqrt{\frac{e^{2}}{m_{p}}\Lambda^{\frac{1}{2}}}
\end{equation}.

Therefore the light cone contains a new region of causality called {\it dark cone}\cite{Nassif}, so that the speed of a particle
must belong to the following range: $V$(dark cone)$<v<c$ (light cone) (Fig.1). 

The breaking of Lorentz symmetry group destroys the properties of the transformations of Special Relativity (SR) and so generates
an intriguing kinematics and dynamics for speeds very close to the minimum speed $V$, i.e., for $v\rightarrow V$, we find new 
relativistic effects such as the contraction of the improper time and the dilation of space\cite{Nassif}. In this new scenario,
the proper time also suffers relativistic effects such as its own dilation with respect to the improper one when 
$v\rightarrow V$, namely:

\begin{equation}
\Delta\tau\sqrt{1-\frac{V^{2}}{v^{2}}}=\Delta t\sqrt{1-\frac{v^{2}}{c^{2}}},
\end{equation}
which was shown in the reference\cite{Nassif}, where it was also made experimental prospects for detecting such new relativistic 
effect close to the invariant minimum speed $V$, i.e., too close to the absolute zero temperature. 

Since the minimum speed $V$ is an invariant quantity as the speed of light $c$, $V$ does not alter the value of the speed $v$ of
any particle. Therefore we have called ultra-referential $S_{V}$\cite{Nassif}\cite{Nassif2} as being the preferred (background) reference
frame in relation to which we have the speeds $v$ of any particle. In view of this, the reference frame transformations change 
substantially in the presence of $S_V$, as follows: 

a) The special case of $(1+1)D$ transformations in SSR\cite{Nassif}\cite{Nassif1}\cite{Nassif2}\cite{Nassif3}\cite{Nassif4} 
with $\vec v=v_x=v$ (Fig.2) are 

\begin{equation}
x^{\prime}=\Psi(x-vt+Vt)=\theta\gamma(x-vt+Vt) 
\end{equation}

and 

\begin{equation}
t^{\prime}=\Psi\left(t-\frac{vx}{c^2}+\frac{Vx}{c^2}\right)=\theta\gamma\left(t-\frac{vx}{c^2}+\frac{Vx}{c^2}\right), 
\end{equation}
where $\theta=\sqrt{1-V^2/v^2}$ and $\Psi=\theta\gamma=\sqrt{1-V^2/v^2}/\sqrt{1-v^2/c^2}$.

\begin{figure}
\includegraphics[scale=0.70]{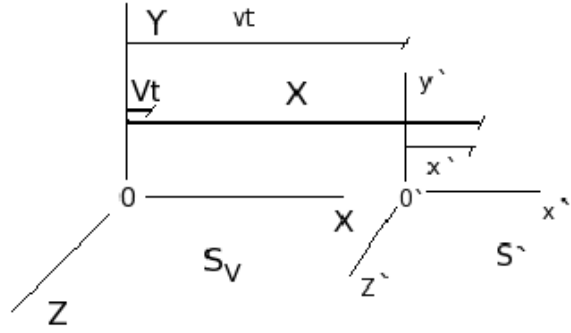}
\caption{In this special case of $(1+1)D$, the referential $S^{\prime}$ moves in $x$-direction with a speed $v(>V)$ with respect to the
 background field connected to the ultra-referential $S_V$. If $V\rightarrow 0$, $S_V$ is eliminated (empty space), and thus the galilean
 frame $S$ takes place, recovering Lorentz transformations.}
\end{figure}

b) The $(3+1)D$ transformations in SSR (Fig.3)\cite{Nassif} are

\begin{equation}
\vec{r'}=\theta\left[\vec{r_{T}}+\gamma\left(\vec{r_{//}}-\vec{v}\left(1-\frac{V}{v}\right)t\right)\right]=
\theta\left[\vec{r_{T}}+\gamma\left(\vec{r_{//}}-\vec{v}t+\vec{V}t\right)\right]
\end{equation}

and

\begin{equation}
t'=\theta\gamma\left[t-\frac{\vec{r}\cdotp\vec{v}}{c^{2}}+\frac{\vec{r}\cdotp\vec{V}}{c^{2}}\right]. 
\end{equation}

Of course, if we make $V\rightarrow 0$, we recover the well-known Lorentz transformations.

\begin{figure}
\includegraphics[scale=0.70]{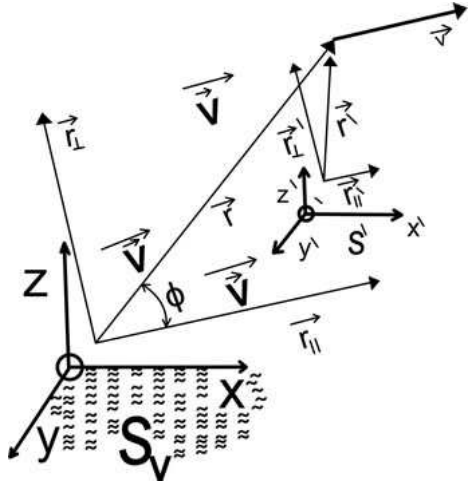}
\caption{$S^{\prime}$ moves with a $3D$-velocity $\vec v=(v_x,v_y,v_z)$ in relation to $S_V$. For the special case of $1D$-velocity
$\vec v=(v_x)$, we recover Fig.2; however, in this general case of $3D$-velocity $\vec v$, there must be a background vector $\vec V$
(minimum velocity) with the same direction of $\vec v$ as shown in this figure. Such a background vector $\vec V=(V/v)\vec v$ is 
related to the background reference frame (ultra-referential) $S_V$, thus leading to Lorentz violation. The modulus of $\vec V$ is invariant
at any direction.} 
\end{figure}

Although we associate the minimum speed $V$ with the ultra-referential $S_{V}$, this frame is inaccessible for any particle. Thus, the
effect of such new causal structure of spacetime generates an effect on mass-energy being symmetrical to what happens close to the speed 
of light $c$, i.e., it was shown that $E=m_0c^2\Psi(v)=m_0c^2\sqrt{1-V^2/v^2}/\sqrt{1-v^2/c^2}$, so that $E\rightarrow 0$ when
 $v\rightarrow V$\cite{Nassif}\cite{Nassif2}. We notice that $E=E_0=m_0c^2$ for $v=v_0=\sqrt{cV}$\cite{Nassif}. It was also shown that 
the minimum speed $V$ is associated with the cosmological constant, which is equivalent to a fluid (vacuum energy) with negative 
pressure\cite{Nassif}\cite{Nassif2}.

The metric of such symmetrical spacetime of SSR is a deformed Minkowski metric with a global multiplicative function (a scale factor
with $v$-dependence) $\Theta(v)$ working like a conformal factor, which leads us to an extended DS-metric to be investigated 
in Sections 4 and 5. Thus we write 

\begin{equation}
ds^{2}=\Theta\eta_{\mu\nu}dx^{\mu}dx^{\nu},
\end{equation}
where $\Theta=\Theta(v)=\theta^{-2}=1/(1-V^2/v^2)$ and $\eta_{\mu\nu}$ is the Minkowski metric.

We can say that SSR geometrizes the quantum phenomena as investigated before (the origin of the Uncertainty Principle)\cite{Nassif3} 
in order to allow us to associate quantities belonging to the microscopic world with a new geometric structure that originates from
Lorentz symmetry breaking. Such a geometry should be investigated in the future. 

\subsection{SSR-metric as a conformal structure}

 Let us consider a conformal transformation $\omega^2$, which is a scale transformation on a metric given in the metric space, thus
leading to a new metric connected to another metric space, namely $\mathcal{G}_{\mu\nu}=\omega^{2}{g}_{\mu\nu}$, which must obey the 
following conditions: 
 
i) $\omega^2$ presents inverse, i.e., $\omega^{-2}$, being a soft function, which generetes all its derivative functions. 
It transforms the neighborhoods of a point $P$, in the topological sense, leading to the neighborhood of a point $P^{\prime}$ 
in the transformed metric space. This is equivalent to isometry properties.

ii) The transformation $\omega^2$ preserves the zero geodesic, namely: 

\begin{equation}
 g_{\mu\nu}X^{\mu}X^{\nu}=g_{\mu\nu}(\omega{x^{\mu}})(\omega{x^{\nu}})=\omega^{2}g_{\mu\nu}x^{\mu}x^{\nu}=
\mathcal{G}_{\mu\nu}x^{\mu}x^{\nu}=0, 
\end{equation}
which means that the transformation $X^{\mu}=\omega{x^{\mu}}$ preserves the space, time and light-like vectors, which also
implies that $\omega>0$. 

iii) It must obey the conservation of angles, namely:

\begin{equation}
 \frac{1}{\sqrt{|w||u|}}g_{\mu\nu}w^{\mu}u^{\nu}=\frac{1}{\sqrt{|W||U|}}\mathcal{G}_{\mu\nu}W^{\mu}U^{\nu}, 
\end{equation}
where $\vec w$, $\vec u$, $\vec W$ and $\vec U$ are any vectors. 

iv) We should consider $\omega(a)$ such that the parameter $a$ needs to belong to the original metric space. 

In order to verify the conditions above on the scale transformation of SSR, where $\mathcal{G}_{\mu\nu}=\Theta(v)\eta_{\mu\nu}$, we
see that $\Theta(v)=\frac{1}{\left(1-\frac{V^2}{v^2}\right)}$ has inverse, so that we find $\Theta^{-1}=\left(1-\frac{V^2}{v^2}\right)$.
This function is defined in the interval $V<v<c$ such that $0<\Theta^{-1}<1-\xi^{2}$, its inverse being in the interval
$\frac{1}{1-\xi^{2}}<\Theta(v)<\infty$, where the speed $v$ is given with respect to the ultra-referential $S_V$\cite{Nassif}. 

In order to realize that $v$ is defined in the metric space of SSR, we should have in mind that $v$ is directly related to measurable 
greatnesses like energy given in the dispersion relation, as follows: 

\begin{equation}
 \Theta\eta_{\mu\nu}p^{\mu}p^{\nu}=m_{0}^{2}c^{2},  
\end{equation}
from where we get

\begin{equation}
 p_{\mu}p^{\mu}=m_{0}^{2}c^{2}\left(1-\frac{V^2}{v^2}\right). 
\end{equation}

Now, verifying the null geodesic, i.e., 

\begin{equation}
 \eta_{\mu\nu}x^{\mu}x^{\nu}=0, 
\end{equation}
and having $\theta=\sqrt{\left(1-\frac{V^2}{v^2}\right)}$, so we find $X^{\mu}=\sqrt{(1-\frac{V^2}{v^2})}x^{\mu}$. Thus, by
applying such transformation to the null geodesic, we obtain 

\begin{equation}
 \eta_{\mu\nu}\sqrt{\left(1-\frac{V^2}{v^2}\right)}x^{\mu}\sqrt{\left(1-\frac{V^2}{v^2}\right)}x^{\nu}=
\left(1-\frac{V^2}{v^2}\right)\eta_{\mu\nu}x^{\mu}x^{\nu}=0. 
\end{equation}

Of course the angle conservation is obeyed by previous construction. In SR we have the well-known {\it boosts}, but in SSR, the 
``boosts'' in the vicinity of the minimum speed ($v{\approx}V$) are deformed in the following way: Consider the symmetric matrix
${\theta}I$\cite{Nassif}, so that a vector in the interval $ds$ becomes invariant under the scale transformation, as follows: 

\begin{equation} 
 x^{*\mu}={\theta}Ix^{\mu}=\sqrt{\left(1-\frac{V^2}{v^2}\right)}Ix^{\mu}, 
\end{equation}
 where $\theta=\Theta^{-1/2}$ is also a scale factor. Alternatively we have written 
$\Lambda(v{\approx}V)x^{\mu}=x^{*\mu}$\cite{Nassif}. 

 We write

  \begin{equation}
   (ds^{*})^{2}=ds^{2}(v)=dx^{*\mu}dx^{*}_{\mu}={\theta}ds^{2}, 
  \end{equation}
where $ds^{2}$ is the squared interval of the usual space-time in SR and $(ds^{*})^{2}$ is the deformed interval that dilates
drastically close to the minimum speed. Thus the effect of the introduction of a minimum speed with the presence of the
ultra-referential $S_{V}$ leads to a scale deformation, so that the deformed interval $(ds^{*})^{2}$ remains invariant. In view of
this, we write 

\begin{equation}
 (ds'*)^{2}=(ds*)^{2}=\eta_{\mu\nu}dx*^{\mu}dx*^{\nu}, 
\end{equation}
where $(ds*)^{2}=(1-\alpha)^{2}dx^{\mu}dx_{\mu}=\theta^{2}ds^{2}$. Of course in the limit of $V\rightarrow 0$ ($\alpha\rightarrow0$), 
the effect of scale transformation vanishes and so SSR recovers SR-theory. 

Now we write SSR-metric, as follows: 

\begin{equation}
 ds^{2}={\Theta}\eta_{\mu\nu}dx^{\mu}dx^{\nu}, 
\end{equation}
where

\begin{equation}
 \Theta=\Theta(v)=\frac{1}{(1-\frac{V^2}{v^2})}. 
\end{equation}

Thus the covariant SSR-metric is under the following scale transformation:  

\begin{equation}
 \mathcal G_{\mu\nu}=\Theta\eta_{\mu\nu}=\frac{1}{\left(1-\frac{V^2}{v^2}\right)}\left[\begin{array}{rrcccccccc}
-1 & 0    & 0  & 0 \\
 0   & 1 & 0    & 0 \\
 0   & 0 & 1 & 0 \\
 0   & 0 & 0 & 1 \\
\end{array} \right]. 
\end{equation}

It is easy to verify that we recover Minkowski metric ($\eta_{\mu\nu}$) in the limit $V\rightarrow 0$. Thus we realize that the
scale transformation of SSR ($\Theta$) generates a conformally flat metric, i.e., $\Theta\eta_{\mu\nu}$. 

\section{Calculating the curvature in dS-relativity}

 Let us begin by writing Einstein equation, namely: 

\begin{equation}
 R_{\mu\nu}-\frac{Rg_{\mu\nu}}{2}=\frac{8{\pi}G}{c^2}T_{\mu\nu}. 
\end{equation}

The energy-momentum tensor is

\begin{equation}
 T_{\mu\nu}=(\rho+p)U_{\mu}U_{\nu}+pg_{\mu\nu}. 
\end{equation}

By making $p=-\rho$, which is the equation of state (EOS) for a relativistic perfect fluid\cite{Sean} representing the dark energy,
we have 

\begin{equation}
 T_{\mu\nu}=-\rho g_{\mu\nu}=pg_{\mu\nu},
\end{equation}
where we must have $p<0$ in order to get $\rho>0$. 

By substituting the equation above in Einstein equation, we find 

\begin{equation}
 R_{\mu\nu}-\frac{Rg_{\mu\nu}}{2}=\frac{8\pi G}{c^2}pg_{\mu\nu}. 
\end{equation}

Manipulating Einstein equation, we obtain

\begin{equation}
 R_{\mu\nu}=\left(\frac{R}{2}+\frac{8{\pi}G}{c^2}p\right)g_{\mu\nu}, 
\end{equation}
where we find the Ricci tensor as being proportional to the metric. Thus we have a maximal space\cite{Sitter1}, which represents 
a space with a non-null constant of curvature $R$. 

By applying the contravariant metric $g^{\mu\nu}$ at both sides of Eq.(25), we get 

\begin{equation}
 R=\frac{16{\pi}G}{c^2}p. 
\end{equation}

By using EOS for vacuum ($p=-\rho$) in Eq.(26), we find 

\begin{equation}
 R=-\frac{16\pi G}{c^2}\rho. 
\end{equation}

We realize that the curvature $R$ is constant and negative as occurs in dS-relativity with $\Lambda>0$\cite{Sitter1, Sitter2, Andersson}.

Taking into account the relationship between the vacuum energy density and the cosmological constant, namely: 

\begin{equation}
 \rho=\frac{{\Lambda}c^2}{8{\pi}G}, 
\end{equation}
and substituting Eq.(28) into Eq.(27), we find the relationship between the curvature and the cosmological constant, as follows:

\begin{equation}
 R=-2\Lambda, 
\end{equation}
where $\Lambda>0$, as occurs in dS-relativity, $R$ being negative.  

\section{Curvatures in SSR-space}

 In SSR, the covariant energy-momentum tensor of a perfect fluid without anisotropy is defined in the following way: 

\begin{equation}
 \mathcal T_{\mu\nu}=(\rho+p)\mathcal U_{\mu}\mathcal U_{\nu}+p\mathcal G_{\mu\nu}, 
\end{equation}
where $\mathcal G_{\mu\nu}=\Theta g_{\mu\nu}$, and $\Theta=\Theta(v)=1/\left(1-\frac{V^2}{v^2}\right)$. We
have $\mathcal U_{\mu}=\left[-\frac{\sqrt{1-\frac{V^2}{v^2}}}{\sqrt{1-\frac{v^2}{c^2}}}~ , ~
\frac{v_{i}\sqrt{1-\frac{V^2}{v^2}}}{c\sqrt{1-\frac{v^2}{c^2}}}\right]$, with $i=1,2,3$. As $\mathcal T_{\mu\nu}$ represents 
a perfect fluid, $\mathcal U_{\mu}$ being the quadri-velocity of the fluid, where a background field is considered for representing
the ultra-referential $S_V$, then $v$ should be interpreted as the velocity of a virtual proof particle in the fluid, moving with respect to
$S_V$ so that, when $v\rightarrow V$, such a proof particle becomes too close to the background frame $S_V$, i.e., a ultra-cold fluid
that works like a ultra-cold gas that emerges in this kind of DS-scenario for a very old and cold expanding universe. So we could say that the cosmological
vacuum in a very old universe is mimicked by a very cold gas so that its quadri-velocity is close to zero, i.e.,
 $\mathcal U_{\mu}\approx 0$, thus leading to the approximation, namely $\mathcal T_{\mu\nu}\approx p\mathcal G_{\mu\nu}$. This result
will be investigated soon.      
  
In a maximal space of Relativity, it is well-known that the energy momentum tensor is proprortional to the metric
\cite{Sitter1, Sean, Sean1}, namely: 

\begin{equation}
 T_{\mu\nu}=Ag_{\mu\nu}
\end{equation}
where $A$ is a real constant\cite{Sean, Sean1}, thus leading to EOS $p=-\rho$, which is associated with 
the dark energy (vacuum energy), so that the dark energy is described as a maximal space that presents a constant curvature $R$.
It is important to stress that the vacuum as a maximal space emerges naturally from $\mathcal T_{\mu\nu}$ of SSR in the
limit of lower energies, so that we find $\mathcal T_{\mu\nu}\approx p\mathcal G_{\mu\nu}$ from Eq.(30). This does not occur with 
$T_{\mu\nu}$ in Relativity, since its quadri-velocity $U_{\mu}$ cannot be nullified, even for $v=0$. 

When we deal with dS-relativity, the dark energy is described by the energy-momentum tensor, as follows: 

\begin{equation}
 T_{\mu\nu}=p\eta_{\mu\nu}=-{\rho}\eta_{\mu\nu}, 
\end{equation}
where $p=-\rho$, $\eta_{\mu\nu}$ being the Minkowski metric. So we write Einstein equation for the dark energy, namely: 

\begin{equation}
 R_{\mu\nu}-\frac{R\eta_{\mu\nu}}{2}=-\frac{8{\pi}G}{c^2}\rho\eta_{\mu\nu}.
\end{equation}

By applying the contravariant metric tensor $\eta^{\mu\nu}$ at both sides of Eq.(33) and reminding 
that $\eta^{\mu\nu}\eta_{\mu\nu}=1$, we find 

\begin{equation}
 R=-\frac{16\pi G}{c^2}\rho, 
\end{equation}
which occurs in dS-relativity that presents a maximal space. So, by substituting Eq.(34) in Eq.(33), we obtain $R_{\mu\nu}$ given 
in dS-relativity, as follows: 

\begin{equation}
 R_{\mu\nu}=-\frac{16{\pi}G}{c^2}\rho\eta_{\mu\nu}, 
\end{equation}
where $\rho$ is the vacuum energy density. 

We already know that SSR-metric ($\mathcal{G}_{\mu\nu}$) is related to the Minkowski metric in the following way: 

\begin{equation}
 \mathcal{G}_{\mu\nu}=\Theta(v)\eta_{\mu\nu}, 
\end{equation}
where SR-metric (Minkowski metric $\eta_{\mu\nu}$) represents a flat space, i.e., with null curvature $R=0$. 

In a maximal space of SSR, by considering $\rho=-p$, we deform the covariant energy-momentum tensor (Eq.32) as follows: 

\begin{equation}
 \mathcal{T}_{\mu\nu}=\Theta T_{\mu\nu}=-\Theta(v)\rho\eta_{\mu\nu}=-\frac{1}{\left(1-\frac{V^2}{v^2}\right)}\rho\eta_{\mu\nu},  
\end{equation}
where $\Theta\eta_{\mu\nu}=\mathcal G_{\mu\nu}$.

Thus, in SSR-scenario, Einstein equation assumes the covariant form, namely:

\begin{equation}
 \mathcal{R}_{\mu\nu}-\frac{\mathcal{R}\eta_{\mu\nu}}{2}=-\frac{8{\pi}G}{c^2}\Theta\rho\eta_{\mu\nu}, 
\end{equation}
where, in Eq.(38) we have considered the effect of cosmological constant at its right side (see Eq.(30) for 
$\mathcal U_{\mu}\approx 0$). 

By applying the contravariant metric $\eta^{\mu\nu}$ at both sides of Eq.(38), we find

\begin{equation}
 \mathcal{R}=-\frac{16\pi G}{c^2}\Theta\rho=-\frac{16\pi G}{c^2}\frac{1}{\left(1-\frac{V^2}{v^2}\right)}\rho. 
\end{equation}

 This result above (Eq.(39)), which leads to an extended dS-relativity, impplies in the changes of symmetry like symmetry groups by
breaking down Lorentz symmetry and the introduction of a conformal  
symmetry for changing of scales with the factor $\Theta$. So we interpret that the introduction of the 
ultra-referential $S_V$ can introduce larger (negative) curvatures, being associated with Mach principle within a quantum scenario where
rest is forbidden (a zero-point energy), since a particle is always interacting with everything around it due to the existence of the 
background field (vacuum energy) represented by $S_V$. 

By substituting Eq.(39) in Eq.(38) and performing the calculations, we find 

\begin{equation}
 \mathcal{R_{\mu\nu}}=\Theta R_{\mu\nu}=-\frac{16\pi G}{c^2}\Theta\rho\eta_{\mu\nu}=
-\frac{16\pi G}{c^2}\frac{1}{\left(1-\frac{V^2}{v^2}\right)}\rho\eta_{\mu\nu},  
\end{equation}
where we see that $R_{\mu\nu}=-(16\pi G/c^2)\rho\eta_{\mu\nu}$ (Eq.(35)) is given in dS-relativity whereas 
$\mathcal{R_{\mu\nu}}=\Theta R_{\mu\nu}$ (Eq.(40)) is given in an extended dS-relativity (SSR). 

 dS-theories consider that the cosmological constant is related to a dS-length\cite{Sitter1, Sitter2}. In view of this, it is
interesting to notice that Nassif\cite{Nassif, Nassif1, Nassif2} has introduced a relationship between the cosmological constant and the 
universe radius $r_{u}$ (the Hubble radius), namely:

 \begin{equation}
 \Lambda=\frac{6c^2}{r^{2}_{u}},
 \end{equation}
where, being $r_u\sim 10^{26}$m, we find $\Lambda\sim 10^{-35}s^{-2}$ in agreement with the observational 
data\cite{Perlmutter}. 

Now we can relate the cosmological constant $\Lambda$ to Ricci curvature $\mathcal R$ in SSR-scenario. To do that, by considering
$\Lambda=8\pi G\rho/c^2$ and also combining Eq.(39) with Eq.(41), we find  

\begin{equation}
\mathcal R_{\Lambda}=\mathcal R=R\Theta=-2\Lambda\Theta=-\frac{16{\pi}G}{c^2}\Theta\rho=
-\frac{16\pi G}{c^2}\frac{1}{\left(1-\frac{V^2}{v^2}\right)}\rho
=-\frac{12c^2}{r^{2}_{u}}\Theta,
\end{equation}
from where we get the curvature of SSR, i.e., $\mathcal R=\Theta R$, $R$ being the curvature of a DS-space\cite{Sean}. 

We have verified that Ricci curvature obtained in the scenario of SSR ($\mathcal{R_{\mu\nu}}$) plays the role of a set of infinite
values of negative curvature, so that we have an extended dS-relativity (Eq.(42)).  

 We realize that, when the universe radius $r_u$ increases to infinite, the scalar curvature of the cosmological vacuum
($\mathcal R_{\Lambda}$) tends to an infinitely negative value. Beside this, the universe becomes colder since the increase of its radius ($r_u$) 
leads to the decrease of velocity $v$ of the fluid mimicked by a gas of virtual particles. When $v\rightarrow V$, we find the fundamental
state of vacuum whose potential becomes the minimum value $-c^2$ as we will show in Section 6 (Fig.4). Thus, in spite of there is
a relationship between the variables $r_u$ and $v$ for obtaining the curvature $\mathcal R_{\Lambda}$, both variables are treated 
separately. But, in Section 6, we intend to make a connection between such variables, since $v$ can also be interpreted as being 
an escape velocity, but $v$ should represent an input speed just in the case of negative (repulsive) gravitational potential connected to 
the radius $r_u$ of a sphere filled by dark energy that expands for representing a spherical dark universe with Hubble radius (Section 6). 

 In Eq.(42), we should note that the vacuum curvature tends to diverges, i.e., $\mathcal R\rightarrow-\infty$ when $v\rightarrow V$,
which means we would have a highly repulsive vaccuum. Beside this, for $r_u\rightarrow\infty$ when $v\rightarrow V$, we would have a
scenario for too long time, so that the universe recovers a rapid increase of acceleration of expansion. This seems to lead to the
so-called ``Big Rip'' when we have $v\rightarrow V$ (ultra-cold universe with $\mathcal R\rightarrow-\infty$), but this question 
deserves to be deeply explored elsewhere.

Here we should call attention to the fact that, as SSR was built in a quasi absence of (attractive) matter in space, i.e., $T_{\mu\nu}\approx 0$, we
have introduced the conception of a quasi-flat spacetime\cite{Nassif}. However, in this work, since we only investigate the nature of vacuum
of SSR with respect to its curvature $\mathcal R$, which represents a set of negative curvatures ($-\infty<\mathcal R<0$), we should have
in mind that the vacuum of SSR is conformally flat as occurs in dS-relativity, which has a maximal space with negative curvature.   

At the present time, one observes that the whole universe is flat due to the fact that its positive curvature emerging from the 
visible matter (vm) plus dark matter (dm) cancels its negative curvature of vacuum energy (dark energy) given by Eq.(42). This is so-called
{\it Cosmic Coincidence Problem}\cite{Sean} that establishes that the total curvature of the present universe is $R_{vm+dm}+\mathcal R_{\Lambda}=0$,
its vacuum curvature being $\mathcal R\approx R$, since $\Theta\approx 1$ at the present time when we still have $v>>V$.   

\subsection{SSR-metric as a set of infinite negative curvatures: An extended dS-relativity}

  We have just related the scale factor $\Theta(v)$ to SSR-curvature, thus having dependence on $v$, so that SSR-metric is  
 $\mathcal{G}_{\mu\nu}(v)=\Theta(v)\eta_{\mu\nu}$, where $\Theta(v)=1/(1-V^2/v^2)$. Now by expanding $\Theta(v)$ in series of
$v$, we write

\begin{equation}
 \Theta(v)=1+\frac{V^2}{v^2}+\frac{V^4}{2!v^4}+\frac{V^6}{3!v^6}+\frac{V^8}{4!v^8}+..., 
\end{equation}

or then 

\begin{equation}
 \Theta(v)=\Sigma^{\infty}_{n=0}\frac{1}{n!}\left(\frac{V}{v}\right)^{2n}.
\end{equation}

Thus we can rewrite SSR-metric as 

\begin{equation}
 \mathcal{G}_{\mu\nu}=\Sigma^{\infty}_{n=0}\frac{1}{n!}\left(\frac{V}{v}\right)^{2n}\eta_{\mu\nu}, 
\end{equation}

or then 

\begin{equation}
\mathcal{G}_{\mu\nu}=\eta_{\mu\nu}+\frac{V^2}{v^2}\eta_{\mu\nu}+\frac{V^4}{2!v^4}\eta_{\mu\nu}+\frac{V^6}{3!v^6}\eta_{\mu\nu}
+\frac{V^8}{4!v^8}\eta_{\mu\nu}+..., 
\end{equation}
where we see that the first term (zero order term) of this series is $\mathcal {G}^{n=0}_{\mu\nu}=\eta_{\mu\nu}$, which represents 
exactly Minkowski metric with null curvature. So, we realize that higher order terms lead to very negative curvatures, which tend to
$\mathcal R\rightarrow -\infty$ if we consider all the infinite terms for $v\rightarrow V$. Thus, we can write the 
variable curvature of SSR in the form 

\begin{equation}
\mathcal R=\Theta R=-\frac{16{\pi}G}{c^2}\rho\left(1+\frac{V^2}{v^2}+\frac{V^4}{2!v^4}+\frac{V^6}{3!v^6}+\frac{V^8}{4!v^8}+...\right).  
\end{equation}

As we have an extended maximal space connected to the extended DS-Relativity (SSR), where the extended Ricci tensor
 $\mathcal R_{\mu\nu}(=\Theta R_{\mu\nu})$ is proportional to the extended metric of SSR $\mathcal G_{\mu\nu}(=\Theta\eta_{\mu\nu})$, 
i.e., $\mathcal R_{\mu\nu}=\Theta R_{\mu\nu}=\Theta R\eta_{\mu\nu}=\mathcal R\eta_{\mu\nu}=-\Theta(16{\pi}G/c^2)\rho\eta_{\mu\nu}$,
then from Eq.(47) we obtain

\begin{equation}
\mathcal R_{\mu\nu}=\mathcal R\eta_{\mu\nu}=\Theta R\eta_{\mu\nu}=
-\frac{16{\pi}G}{c^2}\rho\left(1+\frac{V^2}{v^2}+\frac{V^4}{2!v^4}+\frac{V^6}{3!v^6}+\frac{V^8}{4!v^8}+...\right)\eta_{\mu\nu},   
\end{equation}
which is exactly Eq.(40) given in its expanded form of $\Theta(v)$. 

The n-th term of Eq.(46) is $\mathcal G^{n}_{\mu\nu}=(1/n!)(V/v)^{2n}\eta_{\mu\nu}$, so that we can write the n-th term of curvature
as being $\mathcal R(n,v)=-(16\pi G\rho/c^2)(1/n!)(V/v)^{2n}$. 

\section{SSR-metric as a solution of Einstein equation in a dS-scenario}

We intend to verify that SSR-metric ($\mathcal{G}_{\mu\nu}=\Theta\eta_{\mu\nu}$) satisfies Einstein equation given in 
such DS-scenario of SSR. To do that, we first write the usual Einstein equation with the presence of the cosmological constant 
at its left side, since now we want to incorpore the cosmological term into spacetime geometry, i.e.: 

\begin{equation}
 R_{\mu\nu}-\frac{Rg_{\mu\nu}}{2}+{\Lambda}g_{\mu\nu}=\frac{8{\pi}G}{c^2}T_{\mu\nu}, 
\end{equation}
where we must consider $T_{\mu\nu}=0$ (no matter) for DS-relativities\cite{Sean}\cite{Sean1}. 

Now let us admit that the metric $\mathcal G_{\mu\nu}=\Theta\eta_{\mu\nu}$ is a solution of Einstein equation in our 
DS-scenario of SSR, where we must have $\mathcal T_{\mu\nu}=0$. So, in order to verify this hypothesis, we write the
following equation:   
  
\begin{equation}
\mathcal R_{\mu\nu}-\frac{(R\Theta)}{2}\eta_{\mu\nu}+{(\Lambda\Theta)}\eta_{\mu\nu}=0, 
\end{equation}

or then  

\begin{equation}
\mathcal R_{\mu\nu}-\frac{\mathcal R}{2}\eta_{\mu\nu}+{\varLambda}\eta_{\mu\nu}=0. 
\end{equation}

We should call attention to the fact that Einstein equation above (Eq.(50) or Eq.(51)) given in our DS-scenario for the 
metric $\Theta\eta_{\mu\nu}$ must also contain the informations about the extended Ricci tensor $\mathcal R_{\mu\nu}$, the extended
scalar curvature $\mathcal R$ and the effective cosmological ``constant'' $\varLambda=\Theta\Lambda$. 

Manipulating Eq.(50), we find

\begin{equation}
 \mathcal R_{\mu\nu}=\left(\frac{R}{2}-{\Lambda}\right)\Theta\eta_{\mu\nu}.
\end{equation}

As we already know that $\mathcal R_{\mu\nu}=\mathcal R\eta_{\mu\nu}=\Theta R\eta_{\mu\nu}$ (Eq.(40)), then we write Eq.(50), as
follows: 

\begin{equation}
R\Theta\eta_{\mu\nu}=\left(\frac{R}{2}-{\Lambda}\right)\Theta\eta_{\mu\nu},
\end{equation}

from where we get the following equality: 

\begin{equation}
 R=\left(\frac{R}{2}-{\Lambda}\right). 
\end{equation}

Indeed we see that such equality above leads to $R=-2\Lambda$, which is in agreement with Eq.(29) obtained previously
in DS-relativities. Therefore, we verify that $\Theta\eta_{\mu\nu}$ is the solution of Eq.(50), where $R\Theta=\mathcal R$ 
and $\Lambda\Theta=\varLambda$. 

By using the expansion given in Eq.(43) (or Eq.(44)), we can alternatively write Eq.(40) as follows: 

\begin{equation}
 \mathcal R_{\mu\nu}=-2\Lambda\Sigma^{\infty}_{n=0}\frac{1}{n!}\left(\frac{V}{v}\right)^{2n}\eta_{\mu\nu}, 
\end{equation}

or then 

\begin{equation}
 \mathcal R_{\mu\nu}=-2\Lambda\left(\eta_{\mu\nu}+\frac{V^2}{v^2}\eta_{\mu\nu}+
\frac{V^4}{2!v^4}\eta_{\mu\nu}+\frac{V^6}{3!v^6}\eta_{\mu\nu}+\frac{V^8}{4!v^8}\eta_{\mu\nu}+...\right),  
\end{equation}
where we have $\mathcal R_{\mu\nu}=\mathcal R\eta_{\mu\nu}=R\Theta\eta_{\mu\nu}=-2\Lambda\Theta\eta_{\mu\nu}$, so that 
 $\mathcal R=-2\Lambda\Theta=-2\Lambda\Sigma^{\infty}_{n=0}\frac{1}{n!}\left(\frac{V}{v}\right)^{2n}$. 

\subsection{Showing alternatively that SSR-metric is a solution of Einstein equation in a dS-scenario}

Kunhnel and Radmacher\cite{Kunhenel} have shown that the relationship between the metric, the Ricci tensor and its conformal form  
is the following:

\begin{equation}
\tilde{R}_{ab}-R_{ab}=\frac{1}{\Omega^2}[2\Omega\nabla^{2}\Omega+(\Omega\Delta\Omega-3|\nabla\Omega|^{2})g_{ab}], 
\end{equation}
where $\tilde g_{ab}=\omega^2 g_{ab}=\Omega^{-2} g_{ab}$, so that $\omega=\Omega^{-1}$. So we see that the difference between 
the conformal Ricci tensor and its usual form is related to $\nabla\Omega$, $\Delta\Omega$ and $\nabla^{2}\Omega$, which are respectively 
the gradient, the Laplacian and the Hessian of $\Omega$. Such difference between the Ricci tensors is conserved. We still realize
that the difference between such Ricci tensors is proportional to the metrics, namely: 

\begin{equation}
 \tilde{R}_{\mu\nu}-R_{\mu\nu}\propto{g_{\mu\nu}}\propto\tilde g_{\mu\nu}, 
\end{equation}
if we only have 

\begin{equation}
 \nabla^{2}\Omega=(\Delta\Omega)g_{\mu\nu},  
\end{equation}
where $\Delta\Omega=Tr(\nabla^{2}\Omega)$ and $g_{\mu\nu}$ has null curvature. 

 In the case of SSR, the starting metric is $\eta_{\mu\nu}=\left[ \begin{array}{rrcccccccc}
1 & 0    & 0  & 0 \\
 0   & -1 & 0    & 0 \\
 0   & 0 & -1 & 0 \\
 0   & 0 & 0 & -1 \\
\end{array} \right]$, which is an unitary metric. This is a flat metric, that is to say that its Ricci tensor is null, i.e.,
 $R_{\mu\nu}=0$. 

As the Hessian and Laplacian are functions of the second derivatives of the conformal factor, we see that the
condition in Eq.(58) is sastified. Thus the Ricci tensor of SSR is proportional to the own metric, being a maximal space, namely:

\begin{equation}
 \mathcal{R}_{\mu\nu}\propto\mathcal{G}_{\mu\nu}. 
\end{equation}

The spaces whose metrics are proportional to the Ricci tensor are called as maximal spaces\cite{Sean}. There are
three kind of spaces as being maximal spaces, namely the ADS, DS and Minkowski spaces. Nassif\cite{Nassif} was already able to calculate
the value of the cosmological constant by using SSR-theory, which is a theory that is able to generate a perfect fluid by means of 
the energy-momentum tensor, as follows: 

\begin{equation}
 \mathcal{T}_{\mu\nu}=(p+\rho)\mathcal U_{\mu}\mathcal U_{\nu}+p\mathcal{G}_{\mu\nu}. 
\end{equation}

In the regime where $v\approx V$ or then by considering $p=-\rho$, we write 

\begin{equation}
 \mathcal{T}_{\mu\nu}=-\rho\mathcal{G}_{\mu\nu}=-\frac{\Lambda{c^2}}{8{\pi}G}\mathcal{G}_{\mu\nu}=
-\frac{\Theta\Lambda{c^2}}{8{\pi}G}g_{\mu\nu},
\end{equation}
where $\mathcal{G}_{\mu\nu}=\Theta g_{\mu\nu}=\Theta\eta_{\mu\nu}$. 

As we are in a maximal space, then by multiplying Einstein equation by the factor $\Theta$, we obtain

\begin{equation}
 \mathcal R_{\mu\nu}-\frac{R}{2}\mathcal{G}_{\mu\nu}=\frac{8\pi{G}}{c^2}\mathcal T_{\mu\nu}. 
\end{equation}

Substituting Eq.(62) in Eq.(63) above, we find

\begin{equation}
\mathcal{R}_{\mu\nu}-\frac{R}{2}\mathcal{G}_{\mu\nu}+\Lambda\mathcal{G}_{\mu\nu}=0, 
\end{equation}
where, according to Eq.(56), we have $R=-2\Lambda$ so that $\mathcal{R}_{\mu\nu}=-2\Lambda\mathcal{G}_{\mu\nu}$, which is in agreement
with the condition of proportionality in Eq.(60). 

So, Eq.(64) can be simply written as 

\begin{equation}
R_{\mu\nu}-\frac{R}{2}g_{\mu\nu}+\Lambda g_{\mu\nu}=0, 
\end{equation}
which is the Einstein equation with cosmological constant and without sources given for a dS-universe. So, as an example
of what happens with dS-relativity, we conclude that the conformal transformation $\Theta$ introduces a cosmological constant
into Einstein equation. So SSR-metric is a solution of Einstein equation in dS-scenario.  

\section{SSR-metric and dS-metric} 

Let us consider a spherical universe with Hubble radius $r_u$ filled by a uniform vacuum energy density. On the surface of such a
sphere (frontier of the observable universe), the bodies (galaxies) experience an accelerated expansion (anti-gravity)
due to the whole ``dark mass (energy)" of vacuum inside the sphere. So we could think that each galaxy is a proof body interacting with
the big sphere of dark energy (dark universe) like in the simple case of two bodies interaction. However, we need to show that there is 
an anti-gravitational interaction between the 
ordinary proof mass $m_0$ and the big sphere with a ``dark mass" of vacuum ($M$). To do that, let us first start from the 
well-known simple model of a massive proof particle $m_0$ that escapes from a newtonian gravitational potential $\phi$
on the surface of a big sphere of matter, namely $E=m_0c^2(1-v^2/c^2)^{-1/2}\equiv m_0c^2(1+\phi/c^2)$, where $E$ is its relativistic
energy. Here the interval of escape velocity $0\leq v<c$ is associated with the interval of potential $0\leq\phi<\infty$, where we 
stipulate $\phi>0$ to be the attractive (classical) gravitational potential. 

Now we notice that Lorentz symmetry breaking due to the presence of the ultra-referential $S_V$ connected to the dark energy that 
fills the sphere has origin in a non-classical (non-local) aspect of gravity that leads to a repulsive gravitational potential
($\phi<0$). In order to see such an anti-gravity, let us consider the total energy of a proof particle on the surface 
of such a dark sphere according to SSR\cite{Nassif, Nassif1, Nassif2}, namely:

\begin{equation}
E=m_0c^2\frac{\sqrt{1-\frac{V^2}{v^2}}}{\sqrt{1-\frac{v^2}{c^2}}}=
m_0c^2\left(1+\frac{\phi}{c^2}\right),
\end{equation}
from where we obtain
\begin{equation}
\phi=c^2\left[\frac{\sqrt{1-\frac{V^2}{v^2}}}{\sqrt{1-\frac{v^2}{c^2}}}-1\right], 
\end{equation}
where $m_0$ is the mass of the proof particle, $v$ being its input speed or also its escape velocity from the sphere. If the sphere is governed by
vacuum as occurs in the universe as a whole, then $v$ should be understood as an input speed in order to overcome anti-gravity, 
and thus the factor $\sqrt{1-V^2/v^2}$ prevails for determining the potential. However, for a sphere of matter, $v$ is the well-known 
escape velocity, so that the Lorentz factor takes place.  

From the above equation, we observe two regimes of gravitational potential, namely:

\begin{equation}
\phi= \left\{
\begin{array}{ll}
\phi_{Q}:&\mbox{$-c^2<\phi\leq 0$ for $V(=\xi c)< v\leq v_0$},\\\\
  \phi_{att}:&\mbox{$0\leq\phi<\infty$ for $v_0(=\sqrt{\xi}c=\sqrt{cV})\leq v< c$}, 
\end{array}
\right.
\end{equation}
where $v_0=\sqrt{cV}\sim 10^{-3}$m/s\cite{Nassif}. 

$\phi_{att}$ and $\phi_{Q}$ are respectively the attractive (classical) and repulsive (non-classical or quantum)
potentials. We observe that the strongest repulsive potential is $\phi=-c^2$, which is associated with vacuum energy for the
ultra-referential $S_V$ (consider $v=V$ in Eq.(67)) (Fig.4).

\begin{figure}
\includegraphics[scale=0.9]{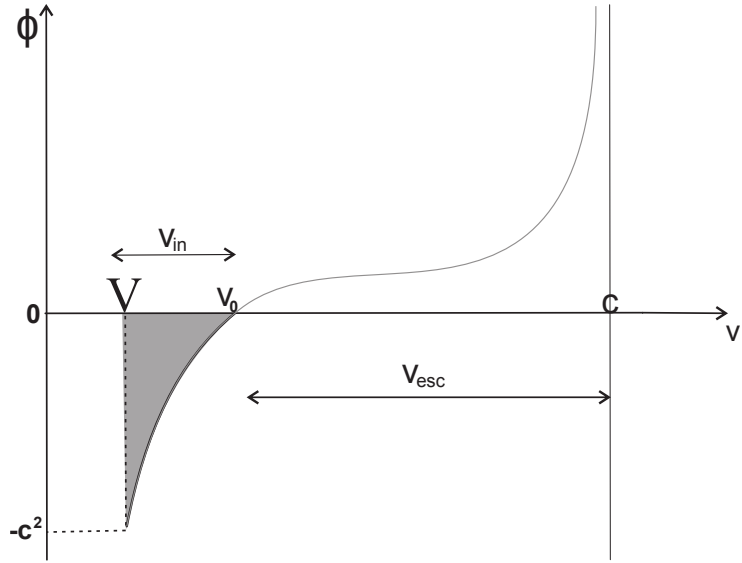}
\caption{This graph shows the potentials of SSR representing the function in Eq.(67) that presents two regimes, namely: a) The attractive (classical)
regime is well-known as Lorentz sector for describing gravity of a source of matter like a sphere of mass having a proof 
particle with mass $m_0$ on its surface. This particle escapes from this gravity with an escape velocity $v_0\leq v_{esc}<c$ according to
the attractive (positive) potential $0\leq\phi_{att}<\infty$. b) The repulsive (quantum) regime is the sector that provides the signature
of SSR for describing anti-gravity of a source of dark energy (vaccum energy) like an exotic sphere of dark mass having a proof 
particle of matter with mass $m_0$ on its surface. In this quantum sector, the escape velocity from anti-gravity should be understood as the 
input velocity $V<v_{in}\leq v_0$ according to the negative (quantum) potential $-c^2<\phi_{Q}\leq 0$, such that the proof particle with mass
$m_0$ is able to penetrate the dark sphere whose anti-gravity pushes it far away. Here we should observe that there is an  
intermediary velocity $v_0=\sqrt{cV}$, which corresponds to the point of phase transition between these two regimes in such a way 
that the general potential $\phi=0$. This means that $v_0$ can represent both escape and input velocities, which depends on the sector
we are considering. As we are just interested in the quantum sector (anti-gravity) of SSR, we have $v_{in}=v_0$ for $\phi=\phi_{Q}=0$ and
$v_{in}=V$ for $\phi=\phi_{Q}=-c^2$, since we just take into account the sector of negative potential for treating the extended
dS-relativity.}
\end{figure}

By considering the simple model of spherical universe with a radius $r_u$ and a uniform vacuum energy
density $\rho$, we find the total vacuum energy inside the sphere, i.e., $E_{\Lambda}=\rho V_u=-pV_u=Mc^2$, 
$V_u$ being its volume and $M$ the total dark mass associated with the dark energy.  Therefore, we are able to get a repulsive 
(negative) gravitational potential $\phi$ on the surface of such a sphere (universe) filled by dark ``mass'' (dark energy), namely: 

\begin{equation}
\phi=-\frac{GM}{r_u}=-\frac{G\rho V_u}{r_uc^2}=\frac{4\pi Gpr_u^2}{3c^2},
\end{equation}
where $p=-\rho$, $\rho$ being the vacuum energy density and $V_u=4\pi r_u^3/3$ (volume of the universe).

Knowing that $\rho=\Lambda c^2/8\pi G$, we write the repulsive potential as follows: 

\begin{equation}
\phi=\phi(\Lambda, r_u)=-\frac{\Lambda r_u^2}{6},
\end{equation}
from where, for any radius $r$ of the expanding universe, we generally write $\phi=-\Lambda r^2/6$\cite{Nassif}. As this potential
represents the repulsive sector of gravity given in Eq.(67), we rewrite Eq.(67) by neglecting the Lorentz factor 
(sector of attractive gravity), and thus we obtain the approximation for the repulsive sector that includes all states of vacuum 
given by $v$ such that $V< v\leq v_0 (=\sqrt{\xi}c)$, namely:  

\begin{equation}
\frac{\phi}{c^2}=-\frac{\Lambda r^2}{6c^2}=\sqrt{1-\frac{V^2}{v^2}}-1, 
\end{equation}
such that, if $v=V$, we find $\phi(V)/c^2=-1$, so that $\phi(V)=-c^2$.  We have $-c^2<\phi<0$ (Fig.4). 

By manipulating Eq.(71), we can rewrite the scale factor $\Theta$ as follows:

\begin{equation}
\Theta=\frac{1}{\left(1-\frac{V^2}{v^2}\right)}=\frac{1}{\left(1+\frac{\phi}{c^2}\right)^2}
=\frac{1}{\left(1-\frac{\Lambda r^2}{6c^2}\right)^2}, 
\end{equation}
where we see that there are three equivalent representations for $\Theta$. 

Substituting Eq.(72) in Eq.(8), we write the spherical metric of SSR in the following way: 

\begin{equation}
d\mathcal S^{2}=-\frac{c^{2}dt^2}{\left(1-\frac{{\Lambda}r^2}{6c^2}\right)^2}+\frac{dr^{2}}{\left(1-\frac{{\Lambda}r^2}{6c^2}\right)^2}
+\frac{r^2(d\theta)^2+r^2\sin^2\theta(d\Phi)^2}{\left(1-\frac{{\Lambda}r^2}{6c^2}\right)^2}, 
\end{equation}
where $r=r_u$. 

We should note that $\phi/c^2=-\Lambda r^2/6c^2$, where we have $-c^2<\phi\leq 0$. So it is interesting to realize that, for the 
approximation $\phi>>-c^2$ or $\left|\phi_{Q}\right|<<c^2$, we are in the regime $v>>V$, which means a weakly repulsive regime that
corresponds to the present time of the universe whose temperature $T(\approx 2.73$K) connected to a certain velocity $v$ is still 
so far from $T=0$K connected to the minimum speed $V$. 

As Eq.(73) encompasses all types of vacuum, specially the ideal vacuum given for a too long time when the universe will be in a very strong
repulsive regime with curvature $\mathcal R\rightarrow -\infty$, now we can realize that only the approximation for a 
weakly repulsive regime is able to generate a special metric well-similar to DS-metric. So, in order to
see this special metric (like DS-metric) given only for weak anti-gravity, we just make the expansion of the denominator
of $\Theta$-factor in Eq.(73), so that we take into account only the first order term, since we are considering $\Lambda r^2/6<<c^2$
such that we have $(1-\Lambda r^2/6c^2)^2\approx (1-2\Lambda r^2/6c^2)=(1-\Lambda r^2/3c^2)$. Finally, in doing this approximation
in SSR-metric [Eq.(73)], we find the DS-conformal metric, namely: 

\begin{equation}
d\mathcal S^{2}\approx dS^{2}_{DS}= -\frac{c^{2}dt^2}{\left(1-\frac{{\Lambda}r^2}{3c^2}\right)}+
\frac{dr^{2}}{\left(1-\frac{{\Lambda}r^2}{3c^2}\right)}+
\frac{r^2d\varOmega}{\left(1-\frac{{\Lambda}r^2}{3c^2}\right)}, 
\end{equation}
where $d\varOmega$ is   

\begin{equation}
 d\varOmega=(d\theta)^2+\sin^2\theta(d\Phi)^2
\end{equation}

We realize that the validity of DS-metric remains only in a weak anti-gravity regime as occurs in the case of the tiny positive 
cosmological constant given in the present time of the expanding universe. 

\section{Relationship between SSR and dS-spacetime}

Here we wish to present the relationship between dS-spacetime and the SSR-spacetime via scalar $\alpha=V/v$, where $V$ is the invariant
minimum speed. 

We can describe dS-spacetime $dS(4,1)$ as being a hyperbolic $4$-surface where the pseudo-Euclidean space 
$\mathbb{E}^{4,1}$ is included by satisfying the following relation\cite{Sitter1}\cite{Pereira1}\cite{Pereira2}: 

\begin{equation}\label{eq1}
\eta_{AB}\zeta^A\zeta^B=\eta_{\alpha\beta}\zeta^{\alpha}\zeta^{\beta}-(\zeta^4)^2= -L^2, 
\end{equation}
where $L$ is the well-known dS-length. 

Now we wish to relate the Euclidean $\zeta^A$ with the stereografic conformal coordinates $x^{\alpha}$ 
$(\alpha, \beta,...=0,1,2,3)$ trhough of the transformations, namely:  

\begin{equation}\label{eq2}
\zeta^{\alpha}=\Omega x^{\alpha};~ ~ \zeta^4=\Omega L,
\end{equation}
where we have admitted that the extra dimension connected to $\zeta^4$ is directly proportional to dS-radius $L$ ($\zeta^4\propto L$),
which will be well justified by SSR-spacetime soon. We have $\Omega=\Omega(x)$ to be determined according to SSR, since the factor 
$\Omega$ should have a direct correspondence with the factor $\Theta$ in SSR. 

The above transformation in dS-space preserves dS-radius $L$, which should have a connection with the invariant minimum speed
$V$ in SSR by means of the factor $\Theta$ and its direct relation with dS-factor $\Omega$ to be investigated. The philosophy of our 
procedure is to compare two invariants that have the same functional origin.
 
Let us write Eq.(77) in the differential form, as follows: 

\begin{equation}\label{eq3}
d\zeta^{\alpha} = x^{\alpha}d\Omega + \Omega dx^{\alpha}; ~ ~ d\zeta^4 = Ld\Omega.
\end{equation}

By introducing the relation $r^2 =\eta_{\alpha\beta}dx^{\alpha}dx^{\beta}$ in Eq.(76), we write

\begin{equation}\label{eq5}
\Omega^2 r^2 -(\zeta^4)^2 = \Omega^2 r^2-\Omega^2 L^2= -L^2, 
\end{equation}
where have considered the simple relation $\zeta^4=\Omega L$ to be consistent with SSR-spacetime as we will justify soon. So
by solving Eq.(79) for determining $\Omega$, we find

\begin{equation}\label{eq6}
\Omega=\Omega(r)=\frac{1}{\sqrt{1-\frac{r^2}{L^2}}}.
\end{equation}

Finally we can write dS-metric in stereografic conformal coordinates, as follows: 

\begin{equation}\label{eq6}
d \Sigma^2 = \eta_{AB} d\zeta^A d\zeta^B = \eta_{\alpha\beta} d\zeta^{\alpha} d\zeta^{\beta} + (d\zeta^4)^2, 
\end{equation}

or then 

\begin{equation}\label{eq7}
d \Sigma^2 = g_{\alpha\beta} dx^{\alpha} dx^{\beta}, 
\end{equation}

so that the conformal metric $g_{\alpha\beta}$ is given by 

\begin{equation}\label{eq8}
g_{\alpha\beta}=\Omega^2\eta_{\alpha \beta} = \frac{1}{\left(1-\frac{r^2}{L^2}\right)}\eta_{\alpha\beta},
\end{equation}
which allows us to relate $\Omega$ with $\Theta$ of SSR. Thus we can perceive that 

\begin{equation}\label{eq9}
\Omega^2=\frac{1}{\left(1-\frac{r^2}{L^2}\right)}\equiv\Theta=\frac{1}{(1-\alpha^2)}, 
\end{equation}
where $\alpha=V/v$. 

As it is known that $L^2=3c^2\Lambda^{-1}$\cite{wang} in DS-space, the equivalence in Eq.(84) is written in the following way:  

\begin{equation}
\Omega^2=\frac{1}{\left(1-\frac{\Lambda r^2}{3c^2}\right)}=\Theta=\frac{1}{\left(1-\frac{V^2}{v^2}\right)}, 
\end{equation}
which represents Eq.(72) given in the approximation of weak anti-gravity (very small cosmological constant), i.e., 
by making $\Lambda r^2/6<<c^2$ in Eq.(72), leading to the approximation $\Theta\equiv(1-\Lambda r^2/6c^2)^{-2}
\approx (1-\Lambda r^2/3c^2)^{-1}=\Omega^2$. Indeed this means that dS-relativity is a special case of SSR when we consider a
weak anti-gravity given in the present time of the universe with a very small value of $\Lambda$\cite{Perlmutter}. In other 
words, we can say that SSR or extended dS-relativity works like a general DS-Relativity, since it is valid for any value of cosmological 
constant, i.e., $0<\Lambda<\infty$. 

It is important to notice that the relationship between the extra dimension $\zeta^4$ and DS-horizon $L$ of DS-space, i.e., 
$\zeta^4=\Omega L$ (Eq.(77)), is in fact consistent with the effect of proper time dilation given close to $V$\cite{Nassif}, since now 
we can write such relation in the form 

\begin{equation}
\zeta^4=\frac{L}{\sqrt{1-\frac{V^2}{v^2}}}\approx\frac{L}{\sqrt{1-\frac{\Lambda r^2}{3c^2}}}, 
\end{equation}
which means that, in the approximation for small values of $\Lambda$ (the current universe), the extra-dimention $\zeta^4$ becomes
slightly larger than $L$, i.e., we have $\Omega=\Theta^{1/2}\approx 1$ ($\zeta^4\approx L$). However, in a very distant future, we expect
that there will be a very strong cosmological anti-gravity as already shown in Section 4. Therefore, for this case ($v\approx V$), we see 
that there should be a dilation of $\zeta^4>>L$, since $\Omega\rightarrow\infty$ for $r\rightarrow L$ (De-Sitter horizon), which means
$v\rightarrow V$. In view of this, we conclude that SSR provides a kinematics foundation for dS-relativity, since 
we can alternatively write Eq.(77) as the proper time dilation in SSR, i.e, $\zeta^4=c\tau=\Theta^{1/2}(ct)$, where $L=ct$. So, 
for a very distant future with $v\rightarrow V$ or a strong anti-gravity $\phi\rightarrow -c^2$ (Eq.(72)), we find that
the proper time in a ``clock'' put close to the De-Sitter horizon elapses much faster than the improper one for a local observer. But
when the proper time is slghtly faster than the improper one ($\zeta^4\approx L$ or $c\tau\approx ct$), we recover the special case
of DS-metric given for weak anti-gravity related to a small cosmological constant. 

We conclude that SSR gives a kinematics explanation for dS-relativity by means of an invariant minimum speed connected to a 
ultra-referential $S_V$, leading to a conformal metric with a varying negative curvature, being able to encompass the DS-metric as a
particular case just for weak anti-gravity (small cosmological constant).

\section{Conclusions and prospects}

We have shown that SSR generates a set of spaces with non-zero (negative) and constant curvatures like maximal spaces. 
We have found that the metric of such spaces (SSR-metric) is a solution of Einstein equation without sources and cosmological
constant. We have verified that the cosmological ``constant'' is in fact a cosmological parameter that depends on the age of the 
universe (its radius), being connected to SSR conformal factor $\Theta(v)$. Actually, we can realize that this scale factor that
generates a deformation on Minkowski metric leads to a new dispersion relation, which reminds Horava theory\cite{Giulia}, however SSR 
takes into account a correction on momentum for infrared regime, i.e., $p^2\approx p_0^2\Theta^{-1}$\cite{Nassif}. In sum, we have 
shown that SSR is a conformally flat theory such as is de Sitter relativity\cite{Sitter4}, so that SSR-metric is a solution of Einstein equation without
matter and with cosmological constant. Therefore SSR introduces the cosmological constant in the Einstein equation as occurs in 
dS-relativity. However, it should be stressed that SSR is not based on a stereographic projection of a fourth component of space, 
but essentially on the existence of a universal minimum speed associated with a thermodynamic background scalar field.

Our perspectives are to search for the development of scalar field models where we associate SSR with the cosmic inflation, including 
the description of cosmic voids phenomena. In the future, we will intend to develop the thermodynamics associated with SSR with the aim 
of studying ideal gas models as raw material for the astrophysical environment. 

We already know that the superfluid generated by SSR is associated with the cosmological constant\cite{Nassif}. In the future, 
such a relationship may allow us to address problems associated with gravitational collapse. The study of the symmetries of SSR must also 
be done soon with the calculation of the symmetries and their association with a new kind of electromagnetism when we have 
$v\rightarrow V$, which could explain the problem of high magnetic fields in magnetars\cite{Greiner}, super-fluids in the interior
of gravastars and other kinds of black hole mimickers (quantum black holes).\\ \\ \\

{\noindent\bf Acknowledgements} \\
Rodrigo Francisco dos Santos is specially greatful to Profa. Maria de F\'atima A. Silva from UERJ (Universidade Estatual do Rio
de Janeiro) and Prof. S\'ergio Ulhoa from UNB (Universidade Nacional de Bras\'ilia) for interesting discussions. He also thanks 
his friend Maur\'icio Moura. The first author C. Nassif is especially greatful to their friends Carlos Magno Leiras, Jorge E. Mattar, 
Em\'ilio C. M. Guerra, J. Alves Correia, Alisson Xavier, Jonas Durval Cremasco, C\'assio Guilherme Reis and Giuseppe Campos Vicentini.
This work is dedicated to the memory of Albert Einstein who searched for the foundation of the cosmological constant abandoned by himself, 
but after it is reborn in a de Sitter scenario based on a universal minimum speed for representing the cosmological vacuum of a new aspect
of quantum gravity within a Machian scenario.


\begin{thebibliography}{99}
\bibitem{Perlmutter} S. Perlmutter et al., Astrophys. J. 517, 565 (1999).
\bibitem{void1} G. W. Gibbons, M. C. Werner, N. Yoshida and S. Chon, arXiv: Astro-ph/13085743. 
\bibitem{void2} M. C. Werner, arXiv: Astro-ph/13117209. 
\bibitem{cosmicinflation} A. Guth, J. Phys.A, 40: 6811-6826 (2007): arXiv:hep-th/0702178.  
\bibitem{Nassif} C. Nassif: {\it On the electrodynamics of moving particlesin a quasi flat spacetime with Lorentz violation and 
its cosmological implications}, International Journal of Modern Physics D Vol.25, 10 (2016) 1650096 (67 pages); \\
see in OPEN ACCESS: http://www.worldscientific.com/worldscinet/ijmpd?journalTabs=read
\bibitem{Nassif1} C. Nassif, Pramana Journal of Physics, Vol.71, 1, p.1-13 (2008). 
\bibitem{Nassif2} C. Nassif: {\it An explanation for the tiny value of the cosmological constant and the low vacuum energy density}, 
General Relativity and Gravitation Vol.47, 9, p.1-34 (2015). 
\bibitem{Nassif3} C. Nassif, {\it Doubly Special Relativity with a minimum speed and the Uncertainty Principle},
International Journal of Modern Physics D, Vol.21, N.2, p.1-20 (2012). 
\bibitem{Nassif4} C. Nassif, {\it Deformed Special Relativity with an energy barrier of a minimum speed},  	
International Journal of Modern Physics D Vol.19, No.5, p.539 (2010). 
\bibitem{Maqueijo} J. Magueijo and L. Smolin, {\it Generalized Lorentz invariance with an invariant energy scale},
Physical Review D.67 (4): 044017 (2003). 
\bibitem{Camelio} G. A. Camelia, {\it Doubly Special Relativity}, Nature 418:34-35 (2002). 
\bibitem{Smolin} J. Magueijo and L. Smolin, {\it Lorentz invariance with an invariant energy scale},
 Physical Review Letters. 88 (19): 190403 (2001). 
\bibitem{Finsler} A. Kostelecky, {\it Riemann-Finsler geometry and Lorentz-violating kinematics}, Phys. Lett.B, 701:137-143 (2011). 
\bibitem{Sitter1} R. Aldrovandi, J. P. Beltr\'an Almeida and J. G. Pereira, {\it De-Sitter special relativity},
Class. Quant. Grav.24:1385-1404, (2007). 
\bibitem{Sitter2} J. P. Beltran Almeida, C. S. O. Mayor and J. G. Pereira, {\it De-Sitter relativity: a natural scenario for an evolving 
$\Lambda$}, Vol.18, 3, p. 181–187 (2012). 
\bibitem{Sitter3} M. Spradlin, A. Strominger and A. Volovich, {\it Les Houches Lectures on De Sitter Space}, arXiv:hep-th/0110007v2.  
\bibitem{Sean} S. Carroll, {\it Spacetime and Geometry: An Introduction to General Relativity-Addison Wesley}, ch.4, 159 (2004).   
\bibitem{Sean1} S. Carroll, Living Rev. Rel.4:1 (2001): arXiv:astro-ph/0004075.    
\bibitem{Andersson} L. Andersson amd G. J. Galloway, Adv. Theor. Math. Phys. 6, p.307–327 (2002). 
\bibitem{Greiner} J. Greiner {\it et al}, Nature 523, 189-192 (2015). 
\bibitem{Giulia}  F. Brighenti, Giulia Gubitosi and J. Magueijo, Phys. Rev.D, 95, 063534 (2017): arXiv:gr-qc/1612.06378 (2016). 
\bibitem{Kunhenel} W. Kuhnel and H. B. Radmacher, Proceeding of the American Mathmatical Society, Vol. 123, n.9, 2841 (1995). 
\bibitem{wang} H. Y. Guo, C. G. Huang and B. Zhou, arXiv: hep-th/0404010.
\bibitem{Sitter4} R. Bousso, arXiv: hep-th/0205177. 
\bibitem{Pereira1} R. Aldrovandi, J. P. Beltr\'an Almeida and J. G. Pereira, arXiv:gr-qc/0403099 (2005). 
\bibitem{Pereira2} A. Araujo, D. F. L\'opez and J. G. Pereira, arXiv:gr-qc/1704.02120 (2017). 
\end{thebibliography}
\end{document}